\documentclass[conference]{IEEEtran}
\IEEEoverridecommandlockouts

\ifCLASSINFOpdf
\else
\fi

\usepackage[T1]{fontenc}  
\usepackage{lmodern}      
\usepackage{times}        

\usepackage{tikz}
\usepackage{pgfplots}
\usepackage{amsmath}
\usepackage{amssymb}
\usepackage{siunitx}
\usepackage{graphicx}
\usepackage{booktabs}
\usepackage{multirow}
\usepackage{caption}
\usepackage{subcaption}

\usepackage[a4paper, left=0.625in, right=0.625in, top=0.75in, bottom=1.7in]{geometry}

\usepackage{caption}
\usepackage{subcaption}
\usepackage{cite}
\usepackage{times}
\usepackage{graphicx}
\usepackage{amsmath}
\usepackage{algorithm}
\usepackage[noend]{algorithmic}
\usepackage{stfloats}
\usepackage{caption}
\usepackage{color}
\usepackage{fancyvrb}
\usepackage{amsmath}
\usepackage{amssymb}
\usepackage{makeidx}
\usepackage{mdwmath}
\usepackage{mdwtab}
\usepackage{moreverb}
\usepackage{fancyvrb}
\usepackage{etoolbox}
\usepackage{mathtools}
\usepackage{mathptmx}
\usepackage{multirow}
\usepackage{color}
\usepackage{url}
\usepackage{tabularx}
\usepackage{enumitem}
\usepackage[bookmarks=false]{hyperref}
\usepackage{footnote}
\usepackage{subcaption}

\makesavenoteenv{figure}

\usepackage{tikz}

\usepackage[nolist]{acronym}

\begin{acronym}[BEREC]
\acro{10G-EPON}{10 Gigabit EPON}
\acro{3GPP}{3rd Generation Partnership Project}
\acro{5G}{Fifth Generation}
\acro{6G}{Sixth Generation}
\acro{A-CPI}{A Controller Plane Interface}
\acro{A-RoF}{Analog Radio-over-Fiber}
\acro{ABNO}{Application-Based Network Operations}
\acro{ADC}{Analog-to-Digital Converter}
\acro{AE}{Allocative Efficiency}
\acro{amc}{adaptive modulation and coding}
\acro{AON}{Active Optical Network}
\acro{AI}{Artificial Intelligence}
\acro{ML}{Machine Learning}
\acro{API}{Application Programming Interface}
\acro{AWG}{Arrayed Waveguide Grating}
\acro{B2B}{Business to Business}
\acro{B2C}{Business to Consumer}
\acro{B-RAS}{Broadband Remote Access Servers}
\acro{BaaS}{Blockchain as a Service}
\acro{BB}{Budget Balance}
\acro{BBF}{BroadBand Forum}
\acro{BBU}{Base Band Unit}
\acro{BER}{Bit Error Rate}
\acro{BEREC}{Body of European Regulators for Electronic Communications}
\acro{BF}{beamforming}
\acro{BFT}{Byzentine fault tolerant}
\acro{BGP}{Border Gateway Protocol}
\acro{BMap}{Bandwidth Map}
\acro{BPM}{Business Process Management}
\acro{BS}{Base Station}
\acro{BSC}{Base Station Controller}
\acro{BtB}{Back to Back}
\acro{BW}{Bandwidth}
\acro{C2C}{Consumer to Consumer}
\acro{C-RAN}{Cloud Radio Access Network}
\acro{C-RoFN}{Cloud-based Radio over optical Fiber Network}
\acro{CA}{Certificate Authority}
\acro{CapEx}{Capital Expenditure}
\acro{CDMA}{code division multiple access}
\acro{CDR}{Clock Data Recovery}
\acro{CFO}{carrier frequency offset}
\acro{CIR}{Committed Information Rate}
\acro{CO}{Central Office}
\acro{CoMP}{Coordinated Multipoint}
\acro{COP}{Control Orchestration Protocol}
\acro{CORD}{Central Office Rearchitected as a Data Centre}
\acro{CPE}{Customer Premises Equipment}
\acro{CPRI}{Common Public Radio Interface}
\acro{CPU}{Central Processing Unit}
\acro{CMC}{Conflict Mitigation Controller}
\acro{CMS}{Conflict Management System}
\acro{CQI}{channel quality indicator}
\acro{CR}{cognitive radio}
\acro{CS}{Central Station}
\acro{CSI}{channel state information}
\acro{D2D}{Device to Device}
\acro{D-CPI}{D Controller Plane Interface}
\acro{D-RoF}{Digital Radio-over-Fiber}
\acro{DAC}{Digital-to-Analog Converter}
\acro{DAS}{distributed antenna system}
\acro{DAS}{Distributed Antenna Systems}
\acro{DBA}{Dynamic Bandwidth Allocation}
\acro{DBRu}{Dynamic Bandwidth Report upstream}
\acro{DC}{Direct Current}
\acro{DL}{Downlink}
\acro{DLT}{Distributed Ledger Technology}
\acro{DMT}{Discrete Multitone}
\acro{DNS}{Domain Name Server}
\acro{DPDK}{Data Plane Development Kit}
\acro{DSB}{Double Side Band}
\acro{DSL}{Digital Subscriber Line}
\acro{DSLAM}{Digital Subscriber Line Access Multiplexer}
\acro{DSP}{Digital Signal Processing}
\acro{DSS}{Distributed Synchronization Service}
\acro{DU}{Distributed Unit}
\acro{DUE}{D2D user equipment}
\acro{DWDM}{Dense Wave Division Multiplexing}
\acro{E-CORD}{Enterprise CORD}
\acro{EAL}{Environment Abstraction Layer}
\acro{EDF}{Erbium-Doped Fiber}
\acro{EFM}{Ethernet in the First Mile}
\acro{EMBS}{Elastic Mobile Broadband Service}
\acro{EON}{Elastic Optical Network}
\acro{EPC}{Evolved Packet Core}
\acro{EPON}{Ethernet Passive Optical Network}
\acro{eMBB}{enhanced Mobile Broadband}
\acro{ETSI}{European Telecommunications Standards Institute}
\acro{EVM}{Error Vector Magnitude}
\acro{FANS}{Fixed Access Network Sharing}
\acro{FASA}{Flexible Access System Architecture}
\acro{FBMC}{filter bank multi-carrier}
\acro{FBMC}{Filterbank Multicarrier}
\acro{FCC}{Federal Communications Commission}
\acro{FDD}{Frequency Division Duplex}
\acro{FDMA}{frequency-division multiple access}
\acro{FEC}{Forward Error Correction}
\acro{FFR}{Fractional Frequency Reuse}
\acro{FFT}{Fast Fourier Transform}
\acro{FPGA}{Field-Programmable Gate Array}
\acro{FSO}{Free-Space-Optics}
\acro{FSR}{Free Spectral Range}
\acro{FTN}{faster-than-nyquist}
\acro{FTTC}{Fiber-to-the-Curb}
\acro{FTTH}{Fiber-to-the-Home}
\acro{FTTx}{Fiber-to-the-x}
\acro{G.Fast}{Fast Access to Subscriber Terminals}
\acro{GEM}{G-PON encapsulation method}
\acro{GFDM}{Generalized Frequency Division Multiplexing}
\acro{GMPLS}{Generalized Multi-Protocol Label Switching}
\acro{GPON}{Gigabit Passive Optical Network}
\acro{GPP}{General Purpose Processor}
\acro{GPRS}{General Packet Radio Service}
\acro{GSM}{Global System for Mobile Communications}
\acro{GTP}{GPRS Tunneling Protocol}
\acro{GUI}{Graphical User Interface}
\acro{HA}{Hardware Accelerator}
\acro{HARQ}{Hybrid-Automatic Repeat Request}
\acro{HD}{half duplex}
\acro{HetNet}{heterogeneous networks}
\acro{IaaS}{Infrastructure as a Service}
\acro{I-CPI}{I Controller Plane Interface}
\acro{I2RS}{Interface 2 Routing System}
\acro{IA}{Interference Alignment}
\acro{IAM}{Identity and Access Management}
\acro{IBFD}{in-band full duplex}
\acro{IC}{Incentive Compatibility}
\acro{ICIC}{inter-cell interference coordination}
\acro{ICT}{information and communications technology}
\acro{IEEE}{Institute of Electrical and Electronics Engineers}
\acro{IETF}{Internet Engineering Task Force}
\acro{IF}{Intermediate Frequency}
\acro{IFFT}{inverse FFT}
\acro{ITS}{Intelligent Transport Systems}
\acro{InP}{Infrastructure Provider}
\acro{IoT}{Internet of Things}
\acro{IP}{Internet Protocol}
\acro{IQ}{In-phase/Quadrature}
\acro{IR}{Individual Rationality}
\acro{IRC}{Interference Rejection Combining}
\acro{ISI}{inter-symbol interference}
\acro{ISP}{Internet Service Provider}
\acro{BFT}{Byzantine-Fault-Tolerant}
\acro{IV}{Intelligent Vehicle}
\acro{ITU}{International Telecommunication Union}
\acro{KPI}{Key Performance Indicator}
\acro{KVM}{Kernel-based Virtual Machine}
\acro{L2}{Layer-2}
\acro{L3}{Layer-3}
\acro{LAN}{Local Area Network}
\acro{LO}{Local Oscillator}
\acro{LOS}{Line Of Sight}
\acro{LR-PON}{Long Reach PON}
\acro{LSP-DB}{Label Switched Path Database}
\acro{LTE-A}{Long Term Evolution Advanced}
\acro{LTE}{Long Term Evolution}
\acro{M-CORD}{Mobile CORD}
\acro{M-MIMO}{massive MIMO}
\acro{M2M}{machine-to-machine}
\acro{MAC}{Medium Access Control}
\acro{ME}{Merging Engine}
\acro{MIMO}{multiple input multiple output}
\acro{MME}{Mobility Management Entity}
\acro{mmWave}{millimeter wave}
\acro{MNO}{mobile network operator}
\acro{MPLS}{Multiprotocol Label Switching}
\acro{MRC}{Maximum Ratio Combining}
\acro{MSC}{Mobile Switching Centre}
\acro{MSP}{Membership Service Providers}
\acro{MSR}{Multi-Stratum Resources}
\acro{MTC}{machine type communication}
\acro{mMTC}{massive Machine Type Communications}
\acro{MVNO}{Mobile Virtual Network Operator}
\acro{MVNP}{mobile virtual network provider}
\acro{MZI}{Mach-Zehnder Interferometer}
\acro{MZM}{Mach-Zehnder Modulator}
\acro{NE}{Network Element}
\acro{NFV}{Network Function Virtualization}
\acro{NFVaaS}{Network Function Virtualization as a Service}
\acro{NG-PON2}{Next-Generation Passive Optical Network 2}
\acro{NGA}{Next Generation Access}
\acro{NLOS}{None Line Of Sight}
\acro{NMS}{Network Management System}
\acro{NRZ}{Non Return-to-Zero}
\acro{NTT}{Nippon Telegraph and Telephone}
\acro{OBPF}{Optical BandPass Filter}
\acro{OBSAI}{Open Base Station Architecture Initiative}
\acro{ODN}{Optical Distribution Network}
\acro{OFC}{Optical Frequency Comb}
\acro{OFDM}{orthogonal frequency-division multiplexing}
\acro{OFDMA}{orthogonal frequency-division multiple access}
\acro{OLO}{Other Licensed Operator}
\acro{OLT}{Optical Line Terminal}
\acro{ONF}{Open Networking Foundation}
\acro{ONU}{Optical Network Unit}
\acro{OOB}{out-of-band}
\acro{OOK}{On-off Keying}
\acro{OpenCord}{Central Office Re-Architected as a Data Center}
\acro{OpEx}{Operating Expenditure}
\acro{OSI}{Open Systems Interconnection}
\acro{OSS}{Operations Support Systems}
\acro{OTT}{Over-the-Top}
\acro{OXM}{OpenFlow Extensible Match}
\acro{P2MP}{Ethernet over point-to-multipoint }
\acro{P2P}{Point-to-Point }
\acro{PAM}{Pulse Amplitude Modulation}
\acro{PAPR}{peak-to-average power rating}
\acro{PAPR}{Peak-to-Average Power Ratio}
\acro{PBMA}{Priority Based Merging Algorithm}
\acro{PCE}{Path Computation Elements}
\acro{PCEP}{Path Computation Element Protocol}
\acro{PCF}{Photonic Crystal Fiber}
\acro{PD}{Photodiode}
\acro{PDCP}{RD Control Protocol}
\acro{PDF}{probability distribution function}
\acro{PGW}{Packet Gateway}
\acro{PHY}{physical Layer}
\acro{PIR}{Peak Information Rate}
\acro{PMD}{Polarization Division Multiplexing}
\acro{PON}{Passive Optical Network}
\acro{POTS}{Plain Old Telephone Service}
\acro{PoET}{Proof of Elapsed Time}
\acro{PoW}{Proof of Work}
\acro{PoS}{Proof of Stake}
\acro{PTP}{Precision Time Protocol}
\acro{PWM}{Pulse Width Modulation}
\acro{QAM}{Quadrature Amplitude Modulation}
\acro{QoE}{Quality of Experience}
\acro{QoS}{Quality of Service}
\acro{QPSK}{Quadrature Phase Shift Keying}
\acro{R-CORD}{Residential CORD}
\acro{RA}{Resource Allocation}
\acro{RAN}{Radio Access Network}
\acro{RAT}{Radio Access Technology}
\acro{RB}{resource block}
\acro{RFIC}{radio frequency integrated circuit}
\acro{RIC}{RAN Intelligent Controller}
\acro{RLC}{Radio Link Control}
\acro{RN}{Remote Node}
\acro{ROADM}{Reconfigurable Optical Add Drop Multiplexer}
\acro{RoF}{Radio-over-Fiber}
\acro{RRC}{Radio Resource Control}
\acro{RRH}{Remote Radio Head}
\acro{RRPH}{Remote Radio and PHY Head}
\acro{RRU}{Remote Radio Unit}
\acro{RSOA}{Reflective Semiconductor Optical Amplifier}
\acro{RU}{Remote Unit}
\acro{Rx}{receiver}
\acro{SC-FDMA}{single carrier frequency division multiple access}
\acro{SCM}{single carrier modulation}
\acro{SD-RAN}{Software Defined Radio Access Network}
\acro{SDAN}{Software Defined Access Network}
\acro{SDMA}{Semi-Distributed Mobility Anchoring}
\acro{SDN}{Software Defined Networking}
\acro{SDR}{Software Defined Radio}
\acro{SFBD}{Single Fiber Bi-Direction}
\acro{SGW}{Serving Gateway}
\acro{SI}{self-interference}
\acro{SIC}{self-interference cancellation}
\acro{SIMO}{Single Input Multiple Output}
\acro{SLA}{Service Level Agreement}
\acro{SMF}{Single Mode Fiber}
\acro{SNMP}{Simple Network Management Protocol}
\acro{SNR}{Signal-to-Noise Ratio}
\acro{SP}{Service Providers}
\acro{Split-PHY}{Split Physical Layer}
\acro{SRI-OV}{Single Root Input/Output Virtualization}
\acro{SU}{secondary user}
\acro{SUT}{System Under Test}
\acro{T-CONT}{Transmission Container}
\acro{TC}{Transmission Convergence}
\acro{TCO}{Total Cost of Ownership}
\acro{TD-LTE}{Time Division LTE}
\acro{TDD}{Time Division Duplex}
\acro{TDM}{Time Division Multiplexing}
\acro{TDMA}{Time Division Multiple Access}
\acro{TED}{Traffic Engineering Database}
\acro{TEID}{Tunnel endpoint identifier}
\acro{TLS}{Transport Layer Security}
\acro{TPS}{Transactions per Second}
\acro{TTI}{transmission time interval}
\acro{TWDM}{Time and Wavelength Division Multiplexing}
\acro{Tx}{transmitter}
\acro{UD-CRAN}{Ultra-Dense Cloud Radio Access Network}
\acro{UDP}{User Datagram Protocol}
\acro{UE}{User Equipment}
\acro{UF-OFDM}{Universally Filtered OFDM}
\acro{UFMC}{Universally Filtered Multicarrier}
\acro{UL}{Uplink}
\acro{UMTS}{Universal Mobile Telecommunications System}
\acro{URLLC}{Ultra-Reliable Low-Latency Communication}
\acro{USRP}{Universal Software Radio Peripheral}
\acro{V2I}{Vehicle to Infrastructure}
\acro{V2V}{Vehicle to Vehicle}
\acro{vBBU}{virtualized BBU}
\acro{vBMap}{Virtual Bandwidth Map}
\acro{vBS}{virtual Base station}
\acro{VCG}{Vickery-Clarke-Groves}
\acro{vCPE}{virtual CPE}
\acro{vCPU}{Virtual Central Processing Unit}
\acro{vDBA}{virtual DBA}
\acro{VDSL2}{Very-high-bit-rate digital subscriber line 2}
\acro{VLAN}{Virtual Local Area Network}
\acro{VM}{Virtual Machine}
\acro{VNF}{Virtual Network Function}
\acro{VNI}{visual networking index}
\acro{VNO}{Virtual Network Operator}
\acro{VNTM}{Virtual Network Topology Manager}
\acro{vOLT}{virtual OLT}
\acro{VPE}{virtual Provider Edge}
\acro{VPN}{Virtual Private Network}
\acro{V2X}{Vehicle to Everything}
\acro{VTN}{Virtual Tenant Network}
\acro{VULA}{Virtual Unblunded Local Access}
\acro{WAN}{Wide Area Network}
\acro{WAP}{Wireless Access Point}
\acro{WCDMA}{wide-band code division multiple access}
\acro{WDM-PON}{Wavelength Division Multiplexing - Passive Optical Network}
\acro{WDM}{Wavelength Division Multiplexing}
\acro{WiMAX}{Worldwide Interoperability for Microwave Access}
\acro{WRPR}{Wired-to-RF Power Ratio}
\acro{XG-PON}{10 Gigabit PON}
\acro{XGEM}{XG-PON encapsulation method}
\acro{XOS}{XaaS Operating System}

\acro{RIC}{RAN Intelligent Controller}
\acro{Near-RT-RIC}{Near Real Time RAN Intelligent Controller}
\acro{Non-RT-RIC}{Non Real Time RAN Intelligent Controller}
\acro{xApp}{Extended Application}
\acro{rApp}{Remote Application}
\acro{CMS}{Conflict Mitigation System}
\acro{EG}{Eisenberg-Galle}
\acro{NSWF}{Nash's Social Welfare Function} 
\acro{CMC}{Conflict Mitigation Controller}
\acro{CDC}{Conflict Detection Controller}
\acro{PMon}{Performance Monitoring}
\acro{RCP}{Recently Changed Parameter}
\acro{PGD}{Parameter Group Definition}
\acro{RCPG}{Recently Changed Parameter Group}
\acro{DCKD}{Decision Correlated with KPI Degradation} 
\acro{KDO}{KPI Degradation Occurrences}
\acro{MNO}{Mobile Network Operator}
\acro{SDL}{Shared Data Layer}
\acro{SON}{Self Organising Network}
\acro{SF}{SON Function}
\acro{AM}{Arithmetic Mean}
\acro{PKR}{Parameter and KPI Ranges}
\acro{KPI}{Key Performance Indicator}
\acro{ICP}{Input Control Parameter}
\acro{RF}{Radio Frequency}
\acro{C-RAN}{Cloud Radio Access Network}
\acro{MARL}{Multi Agent Reinforcement Learning}
\acro{CCO}{Capacity and Coverage Optimization}
\acro{ES}{Energy Saving}
\acro{MRO}{Mobility Robustness Optimization}
\acro{MLB}{Mobility Load Balancing}
\acro{V-RAN}{Virtualized Radio Access Network}
\acro{O-RAN}{Open Radio Access Network}
\acro{D-RAN}{Distributed Radio Access Network}
\acro{5GB}{5G and Beyond}
\acro{O-RU}{Open Radio Unit}
\acro{O-DU}{Open Distribution Unit}
\acro{O-CU}{Open Centralised Unit}
\acro{O-CU-CP}{Open Centralised Unit - Control Plane}
\acro{O-CU-UP}{Open Centralised Unit - User Plane}
\acro{D-SON}{Distributed SON}
\acro{C-SON}{Centralised SON}
\acro{CIO}{Cell Individual Offset}
\acro{CBR}{Call Block Ratio}
\acro{HOR}{Handover Ratio}
\acro{CAN}{Cognitive Autonomous Network}
\acro{CF}{Cognitive Function}
\acro{QACM}{QoS-Aware Conflict Mitigation}
\acro{ANN}{Artificial Neural Network}
\acro{PR}{Polynomial Regression}
\acro{MSE}{Mean Squared Error}
\acro{EVS}{Explained Variance Score}
\acro{CS}{Conflict Supervision}
\acro{TXP}{Tx Power}
\acro{TTT}{Time-To-Trigger}
\acro{CIO}{Cell Individual Offset}
\acro{NL}{Neighbour List}
\acro{RET}{Radio Electrical Tilt}
\acro{HYS}{Handover Hysteresis}
\acro{QACMP}{QACM Priority}
\acro{DRL}{Deep Reinforcement Learning}
\acro{gNB}{Next-generation NodeB}
\acro{RSRP}{Reference Signal Received Power}
\end{acronym}

\usepackage{titlesec}
\titlespacing*{\section}{0pt}{*0.5}{*0.3} 
\titlespacing*{\subsection}{0pt}{*0.4}{*0.2} 

\usepackage{setspace}
\usepackage{enumitem}
\setlist{nosep} 
\setlist[itemize]{left=0pt} 
\setlist[enumerate]{left=0pt}

\begin{document}



\title{RU Energy Modeling for O-RAN in ns3-oran}



\author{\IEEEauthorblockN{Abdul Wadud\IEEEauthorrefmark{1}\IEEEauthorrefmark{2},~\IEEEmembership{Graduate Student Member,~IEEE}
 and} 
\IEEEauthorblockN{Nima Afraz \IEEEauthorrefmark{1}\IEEEauthorrefmark{3},~\IEEEmembership{Senior Member,~IEEE}}
\IEEEauthorblockA{\IEEEauthorrefmark{1}School of Computer Science,
University College Dublin, Ireland}
\IEEEauthorblockA{\IEEEauthorrefmark{2}Bangladesh Institute of Governance and Management, Dhaka, Bangladesh}
\IEEEauthorblockA{\IEEEauthorrefmark{3} UCD Beijing-Dublin International College (BDIC)}
\thanks{Corresponding author: Abdul Wadud (email: abdul.wadud@ucdconnect.ie).}}

%



\IEEEtitleabstractindextext{%
\begin{abstract}
This paper presents a detailed and flexible power consumption model for Radio Units (RUs) in O-RAN using the \texttt{ns3-oran} simulator. This is the first \texttt{ns3-oran} model supporting xApp control to perform the RU power modeling. In contrast to existing frameworks like EARTH or VBS-DRX, the proposed framework is RU-centric and is parameterized by hardware-level features, such as the number of transceivers, the efficiency of the power amplifier, mmWave overheads, and standby behavior. It enables simulation-driven assessment of energy efficiency at various transmit power levels and seamlessly integrates with ns-3's energy tracking system. To help upcoming xApp-driven energy management strategies in O-RAN installations, numerical research validates the model's capacity to represent realistic nonlinear power scaling. It identifies ideal operating points for effective RU behavior.
\end{abstract}

}


\maketitle

\IEEEdisplaynontitleabstractindextext

%
\IEEEpeerreviewmaketitle

\section{Introduction}
\label{sec:intro}

The Open Radio Access Network (O-RAN) encourages disaggregation, interoperability, and intelligent resource management and has completely changed the design paradigm of mobile networks \cite{wadud2023conflict, wadud2024qacm}. The promise of O-RAN to improve energy efficiency is a major factor in its acceptance, as it addresses the environmental impact and rising operational costs of dense 5G installations \cite{oran_potential_2025, rimedo_whitepaper_2023}. A variety of energy-saving techniques have been investigated recently, ranging from intelligent xApps that optimize radio resource allocation based on traffic patterns to dynamic CPU scheduling and base station sleep modes \cite{hoffmann2024energy, intel_vran_2023}. 
Despite these advancements, there is still a sizable gap in open-source simulation platforms like ns-3's ability to estimate energy usage at the \ac{RU} level. Although most research is being done on network-level energy models or high-level abstractions \cite{akman2024energy, wang2023minimizing}, component-specific \ac{RU} energy models that capture dependencies on transmission power, bandwidth, and advanced configurations like mmWave or carrier aggregation is absent. This limitation hampers the ability to evaluate fine-grained energy optimization strategies in realistic O-RAN scenarios.

To fill this gap, a detailed   RU energy consumption model designed specifically for the \texttt{ns3-oran} simulator is presented in this study. Our model allows accurate simulation of RU power usage under various operational scenarios by extending foundational frameworks such as the EARTH model \cite{auer2011energy} to integrate O-RAN-specific elements (e.g., massive MIMO, sleep modes, and dynamic scheduling). We facilitate the shift to 5G-Advanced and upcoming 6G networks by incorporating this model into ns-3, giving researchers and industry practitioners a tool to build and evaluate energy-efficient O-RAN deployments.

The remaining part of the paper is structured as: Section~\ref{sec:background} discusses the state-of-the-art and motivation of this work, Section~\ref{sec:energyModel} details the introduced energy consumption model, Section~\ref{sec:ru-integration} explains the integration of the introduced model in ns-3 framework, Section~\ref{sec:sim_results} illustrates the simulation process and discusses the results, and finally the paper concludes in Section~\ref{sec:conclusion} with concluding remarks.

\section{Background}
\label{sec:background}
The energy efficiency and power consumption modeling of the \ac{RAN} have been extensively explored in recent research. This section mainly discusses the power consumption modeling in the RU in the context of O-RAN based on relevant state-of-the-art studies. 

The EARTH framework \cite{auer2011energy} introduces a holistic approach to energy evaluation of wireless networks, covering various components such as power amplifiers, RF front ends, baseband units, DC-DC converters, and cooling systems. This comprehensive approach is foundational for accurate power estimation in traditional base stations, offering baseline insights applicable to modern O-RAN implementations. \cite{boyapati2010energy} delivers a thorough analysis of energy consumption specifically in LTE baseband subsystems, identifying significant power-consuming functions such as turbo coding, MIMO detection, FFT/IFFT processing, and channel estimation. This model provides precise measurements and guidelines that enhance baseband efficiency.

Liang et al. \cite{liang2024energy} present a detailed analysis of energy consumption specifically targeting O-RAN RUs, but adopting the existing EARTH model which does not accurately fit O-RAN architecture. Their review highlights key RU components, such as RF transceivers, analog-to-digital converters, power amplifiers, and dedicated computing processors (CPUs). They emphasize that these elements significantly contribute to RU energy consumption, particularly due to power amplifier inefficiencies and processing overhead in FPGA or ASIC circuits. Abubakar et al. \cite{abubakar2023energy} presents power consumption models specifically tailored for O-RAN systems. They detail the RU components (RF transceivers, amplifiers, A/D converters, CPUs), emphasizing dynamic function placement impacts and hardware virtualization considerations. Their study also underscores the significant role of transport networks (front-haul, mid-haul, back-haul) in total energy consumption, depending on the functional split implemented. 
The authors in \cite{nafea2021study} explore energy-saving methods for Virtual Base Station (VBS) architectures, which decouple and virtualize baseband and radio operations across cloud infrastructure. Discontinuous Reception (DRX)-based sleep state transitions are used in this model to simulate the power consumption of the VBS and enable components like the RRU and BBU to switch to low-power modes when the system is idle. The modularity and externalized control of RUs in O-RAN installations are not captured by this model, which is useful for simulating large base stations and centralized BBU scenarios but implies integrated scheduling logic.

Despite these developments, the specific architecture and energy behavior of O-RAN RUs cannot be adequately modeled by current models like the EARTH framework and DRX-based VBS methods. The RU/CU/DU components are not separated in the EARTH model, which assumes closely linked monolithic base stations. Similarly, the real-time, xApp-driven control plane and open fronthaul interface of O-RAN RUs are not reflected in DRX-based models, which focus on centralized baseband processing and traffic-aware sleep control. Additionally, these models fail to explicitly represent hardware-level losses such as transceiver count, power supply inefficiencies, and antenna feeder attenuation—factors essential for precise RU-level energy modeling under O-RAN's disaggregated design.

\begin{table*}[h!]
    \centering
    \caption{Comparison of Power Consumption Models for RAN Architectures}
    \label{tab:power-model-comparison}
    \renewcommand{\arraystretch}{1.2}
    \resizebox{\linewidth}{!}{
    \begin{tabular}{|l|c|c|c|}
        \hline
        \textbf{Feature} & \textbf{EARTH Model \cite{auer2011energy}} & \textbf{VBS-DRX Model \cite{nafea2021study}} & \textbf{Introduced RU Model} \\
        \hline
        Target Architecture & Traditional Monolithic BS & Virtual Base Station (VBS) & O-RAN Radio Unit (RU) \\
        \hline
        Control Mechanism & Integrated Baseband Scheduler & DRX Sleep Transitions & External xApp-Driven Control \\
        \hline
        Component Scope & PA, RF, BB, DC/DC, Cooling & RRU, BBU, DRX State Machine & PA, RF, A/D, CPUs, Power Losses \\
        \hline
        RU Specificity & $\times$ (Base station-level) & $\times$ (BBU-centric) & \checkmark (RU-focused) \\
        \hline
        Programmability & $\times$ & $\times$ & \checkmark (O-RAN xApp-compliant) \\
        \hline
        Fronthaul Impact Modeled & $\times$ & $\times$ & \checkmark (Supports disaggregated fronthaul) \\
        \hline
        Transceiver Scaling Support & \checkmark & $\times$ & \checkmark(via parameterized $n_{\mathrm{trx}}$) \\
        \hline
        Hardware Loss Modeling & Basic (Cooling, DC/DC) & DRX Logic Only & \checkmark (PA inefficiency, feeder loss, etc.) \\
        \hline
        Simulation Integration & Offline Analysis & Partial Integration & \checkmark (ns-3 integrated simulation) \\
        \hline
    \end{tabular}
    }
    \vspace{-0.5cm}
\end{table*}

Our introduced framework focuses on the O-RAN RU, a modular hardware unit coupled to the fronthaul, in contrast to other energy models like EARTH and VBS-based DRX methods. Our introduced RU power model breaks down energy consumption into its components and enables simulation-driven assessment of energy efficiency under different transmission activity, transceiver design, and TxPower situations. It provides dynamic control through external xApps and complies with O-RAN's open interface requirements. Table \ref{tab:power-model-comparison} provides a comparison between our model and existing models.


\section{RU Energy Consumption Model for O-RAN}
\label{sec:energyModel}
The RU significantly influences overall power consumption in O-RAN architecture due to specialized components like RF transceivers, power amplifiers, analog-to-digital converters (ADCs), digital-to-analog converters (DACs), and dedicated computational units (CPUs, ASICs, FPGAs) used for real-time physical layer processing tasks such as FFT/IFFT, beamforming, modulation, and filtering \cite{liang2024energy, abubakar2023energy}. In this paper we introduce a comprehensive power consumption model for the RU in \texttt{ns3-oran} \cite{ns3oran_github} that will cover all these aspects. Additionally, to fully capture RU energy behavior, we include \textbf{sleep mode} or \textbf{standby mode} consumption. In sleep or standby mode, the RU disables active transmission/reception chains but maintains minimal functionality for synchronization and monitoring. The sleep power consumption \(P_{\text{RU,sleep}}\) is modeled as: $P_{\text{RU,sleep}} = P_{\text{standby}}$. Hence, the RU power model becomes:



\begin{equation}
P_{\text{RU,total}} = 
\begin{cases}
P_{\text{active}} & \text{(RU active)} \\\\
P_{\text{standby}} & \text{(RU in sleep mode)}
\end{cases}
\label{eq:ru_total_power}
\end{equation}

This allows dynamic RU energy modeling during traffic off-peak periods or adaptive sleep strategies. For active mode, an accurate RU power consumption model, inspired by the EARTH framework \cite{auer2011energy}, can be expressed as:
\begin{equation}
P_{\text{active}} = n_{\text{trx}} \cdot \frac{P_{\text{PA}} + P_{\text{RF}} + P_{\text{BB,proc}} + P_{\text{mmWave}}}{(1 - \delta_{\text{DC}})(1 - \delta_{\text{MS}})(1 - \delta_{\text{cool}})}
\label{eq:ru_active_power}
\end{equation}

However, most power consumption factors except $P_{\text{PA}}$ remains constant which we can represent as $ P_0 = P_{\text{RF}} + P_{\text{BB,proc}} + P_{\text{mmWave}}$. Therefore, the Eq.~\ref{eq:ru_total_power} can be simplified as: 

\begin{equation}
\nonumber
P_{\text{active}} = n_{\text{trx}} \cdot \frac{P_{\text{PA}} + P_0}{(1 - \delta_{\text{DC}})(1 - \delta_{\text{MS}})(1 - \delta_{\text{cool}})}
\label{eq:ru_active_power2}
\end{equation}

In sleep mode, the $P_{\text{PA}} = 0$ as the resting period sets $P_\mathrm{tx}$ to $0$ (see in Eq.~\ref{eq:pa_power}). Therefore, the standby mode power consumption is expressed as: 

\begin{equation}
P_{\text{standby}} = n_{\text{trx}} \cdot P_{\text{sleep}} ~~~~[ \text{where, } P_{\text{sleep}} < P_0]
\label{eq:ru_sleep_power}
\end{equation}

Assuming a DC voltage supply is represented by \(V_{\text{DC}}\), the total RU current \(I_{\text{RU,total}}\) is computed as:

\begin{equation}
I_{\text{RU,total}} = \frac{P_{\text{RU,total}}}{V_{\text{DC}}}
\label{eq:total_ru_current}
\end{equation}

Individual terms used in Eq.~\ref{eq:ru_active_power}-Eq.~\ref{eq:total_ru_current} represent: 

\textbf{Number of Transceivers \(n_{\text{trx}}\)}: Represents the total number of transmit-receive chains in the RU. Each of them are capable of handling a distinct spatial or frequency stream that directly affects overall power consumption.

\textbf{RF Power Amplifier Consumption \(P_{\text{PA}}\)}: The largest energy contributor, directly linked to the transmission power \(P_\mathrm{tx}\) and amplifier efficiency \(\eta_{\text{PA}}\):
    \begin{equation}
    P_{\text{PA}} = \frac{P_\mathrm{tx}}{\eta_{\text{PA}} (1- \delta_{\text{af}})}
    \label{eq:pa_power}
    \end{equation}
Here, \(\eta_{\text{PA}}\) is the efficiency of the power amplifier and $\delta_{\text{af}}$ is the power loss of the antenna feeder. We consider the antenna feeder power loss in Eq.~\ref{eq:pa_power} because the antenna locations of the macro cell is often different from the base station. However, it can be mitigated if we use \ac{RRH}. Typical values of \(\eta_{\text{PA}}\) range between 20\%-40\% and the value of $\delta_{\text{af}}$ is considered to be $-3$ dB \cite{auer2011energy, liang2024energy}. Excluding the feeder power loss, the RF Power Amplifier Consumption \(P_{\text{PA}} =  \frac{P_\mathrm{tx}}{\eta_{\text{PA}}} \), which is ideal for Open \ac{RAN}'s context. 



\textbf{RF Transceiver Consumption \(P_{\text{RF}}\)}: The RF transceiver includes uplink and downlink transmitters and receivers, ADCs, DACs, mixers, and oscillators. It performs signal modulation/demodulation and frequency conversion between analog RF and digital baseband. Most of the power sonsumption in the RF transceiver is associated with the bandwidth demand, the permissible signal-to-noise-and-distortion ratio (SINDR), and the resolution of ADCs and DACs. Typical power consumption ranges between 10–40 W depending on RU complexity and implementation technology (ASIC or FPGA) \cite{liang2024energy, abubakar2023energy}.

\textbf{Baseband Processing \(P_{\text{BB,proc}}\)}: Represents RU-localized baseband processing tasks such as FFT/IFFT, signal detection, channel coding/decoding, beamforming, and all digital signal processing before/after the ADC/DAC stage \cite{boyapati2010energy, liang2024energy}. Its typical consumption ranges from 10–30 W depending on processing intensity and technology used (FPGA/ASIC).

\textbf{RU Power Loss}: In line with the EARTH model \cite{auer2011energy}, supply and cooling inefficiencies are incorporated using multiplicative loss factors. It includes power losses in cooling systems, monitoring circuits, DC-DC converters, and other ancillary components. Approximate losses for each component are listed below:
    \begin{itemize}
        \item \( \delta_{\text{DC}} \): DC-DC conversion loss (typically 5--7\%)
        \item \( \delta_{\text{MS}} \): Mains supply loss (typically 9\%)
        \item \( \delta_{\text{cool}} \): Active cooling loss (typically 10\% for macro cells, 0 otherwise)
    \end{itemize}

\textbf{mmWave-specific Processing \(P_{\text{mmWave}}\)}: This term captures the additional power overhead in mmWave-capable RUs due to extremely high sampling rates, hybrid/digital beamforming, and control logic for large antenna arrays. It includes high-speed digital precoding, RF chain routing, and antenna calibration that are not present in sub-6 GHz RUs. The estimated range is 20–40 W depending on the number of TRX chains, typically modeled as: $P_{\text{mmWave}} \approx (P_{\text{precoding}} + P_{\text{routing}} + P_{\text{calib}})$

For example, in a digital beamforming setup with 64 antennas, assume $P_{\text{precoding}} = 15$~W, $P_{\text{routing}} = 15$~W, and $P_{\text{calib}} = 10$~W. This results in \(P_{\text{mmWave}} \approx 40\)~W, which aligns with expected ranges for mmWave-capable RUs. Hybrid beamforming would lower the routing overhead but still incur analog phase shifting costs. This modular formulation allows $P_{\text{mmWave}}$ to scale with the antenna count and processing architecture, supporting customization for varied RU designs in O-RAN scenarios.

\begin{table*}[h!]
\centering
\caption{Assumed RU Power Model Parameters for Future O-RAN Deployments with Massive MIMO and mmWave}
\label{tab:mmWave_massive_mimo_power}
\resizebox{\linewidth}{!}{
\begin{tabular}{|l|c|c|c|c|c|}
\hline
\textbf{Component} & \textbf{Macro} & \textbf{Micro} & \textbf{Pico} & \textbf{Femto} & \textbf{mmWave Small Cell} \\ 
\hline\hline

Max Tx Power, \(P_\mathrm{tx}^{\text{max}}\) [dBm] & 43--49 & 38--43 & 30--35 & 20--25 & 20--33 \\ 
\hline
Back-off (PAPR margin) [dB] & 9--12 & 9--12 & 8--10 & 6--8 & 8--10 \\ 
\hline
Max PA Output (peak) [dBm] & 58--65 & 50--55 & 38--45 & 26--33 & 30--43 \\ 
\hline
PA Efficiency, \(\eta_{\text{PA}}\) [\%] & 25--35 & 30--40 & 35--45 & 40--50 & 25--40 \\ 
\hline
\textbf{Total PA Consumption}, \(P_{\text{PA}}\) [W] & 200--400 & 40--100 & 10--30 & 1--5 & 5--25 \\ 
\hline
\textbf{RF Transceiver Consumption}, \(P_{\text{RF}}\) [W] & 30--50 & 20--35 & 8--15 & 2--8 & 10--20 \\ 
\hline
\textbf{Baseband Processing}, \(P_{\text{BB,proc}}\) [W] & 100--200 & 50--100 & 20--40 & 5--15 & 15--50 \\ 
\hline
\textbf{mmWave-specific Processing} [W] & -- & -- & -- & -- & 20--40 \\ 
\hline
\textbf{Miscellaneous Overhead}, \(P_{\text{misc,RU}}\) [\%] & 10--20 & 10--15 & 8--12 & 5--10 & 10--15 \\ 
\hline\hline
\textbf{Power per TRX Chain} [W] & 380--750 & 140--295 & 50--105 & 10--30 & 50--135 \\ 
\hline
Number of Antennas (Massive MIMO) & 64--256 & 32--128 & 16--64 & 4--16 & 64--256 \\ 
\hline
Number of TRX Chains & 16--64 & 8--32 & 4--16 & 2--8 & 16--128 \\ 
\hline
\textbf{Total BS Power Consumption} [kW] & 6--20 & 1--4 & 0.2--1 & 0.02--0.24 & 1--6 \\ 
\hline
Approximate Coverage Range & 1--5 km & 200--500 m & 50--200 m & 10--50 m & 50--300 m \\ 
\hline
\end{tabular}
}
\vspace{-0.5cm}
\end{table*}


\subsection{RU Power Model Parameters}
\label{sec:ru-model-params}
Table~\ref{tab:mmWave_massive_mimo_power} summarizes the assumed values for key parameters of a comprehensive power model specifically tailored for future O-RAN deployments that integrate advanced features such as Massive MIMO and mmWave technologies. The parameter values assumed are justified based on existing literature, industry standards, and anticipated technological progressions beyond current 5G deployments.

\textbf{Maximum Transmission Power ($P_\mathrm{tx}^{\text{max}}$)}: The ranges for maximum transmission power reflect typical power levels needed for various cell sizes to ensure appropriate coverage and quality-of-service (QoS). 

\textbf{Back-off (PAPR Margin)}: The Peak-to-Average Power Ratio (PAPR) margin values are selected considering advanced modulation schemes (e.g., OFDM-based Massive MIMO), which inherently have higher peak power requirements. The upper bound is due to higher-order modulation and advanced beamforming, while smaller cells exhibit lower PAPR margins due to less aggressive modulation and fewer antennas.

\textbf{Max PA Output (Peak)}: The peak power amplifier (PA) output directly correlates with the maximum transmission power plus the back-off margin required to accommodate the peak power levels without distortion. Hence, macro cells have the highest PA peak outputs, progressively decreasing through micro, pico, and femto configurations.

\textbf{PA Efficiency (\(\eta_{\text{PA}}\))}: PA efficiency varies significantly with cell size and amplifier technology. Macro cells, typically employing high-power linear PAs, exhibit lower efficiency. Conversely, smaller cells use more efficient amplifier technologies (GaN, CMOS-based PAs), enabling higher efficiencies. However, mmWave small cells experience slightly reduced efficiencies due to technological challenges and higher-frequency operation inefficiencies.

\textbf{Miscellaneous Overhead (\(P_{\text{misc,RU}}\))}: These percentages (5–20\%) account for cooling systems, DC-DC power supply conversions, monitoring electronics, and other infrastructure components. Macro and mmWave small cells, requiring more sophisticated cooling and DC conversion systems due to higher total consumption and component complexity, reflect higher overhead percentages (10–20\%) compared to small indoor cells with minimal cooling demands (5–10\%) \cite{abubakar2023energy}.

\textbf{Power per TRX Chain}: This parameter aggregates the above-discussed consumption elements (PA, RF, baseband, and overhead) to provide a clear view of the total power per transmission and reception (TRX) chain. It scales significantly according to cell complexity and total PA output, ranging from tens of watts (Femto) to hundreds (Macro).

\textbf{Number of Antennas and TRX Chains}: Massive MIMO antenna configurations directly affect computational and power amplifier requirements. Macro and mmWave small cells typically deploy the highest antenna counts (up to 256 antennas), thus substantially influencing the overall power consumption and complexity \cite{liang2024energy}.

\textbf{Total BS Power Consumption}: Provides a comprehensive, realistic estimate of total base station power, reflecting all antenna/TRX chains, overhead, and massive MIMO/mmWave features. The range spans from kilowatts (Macro cells, 6–20 kW) to fractions of kilowatts (Femtocells, 20–240 W), accurately demonstrating the scalability of power needs across cell configurations and types.

\textbf{Approximate Coverage Range}: Reflecting the trade-off between higher transmission power and effective coverage radius, these ranges are based on practical propagation scenarios. Macro cells ensure large-area coverage (1–5 km), while mmWave and small cells provide highly localized coverage (50–300 m and 10–200 m, respectively) due to propagation characteristics at high frequencies \cite{liang2024energy}.

\subsection{Parameter Tuning and Specialization}
Table~\ref{tab:mmWave_massive_mimo_power} contains parameters that represent standard O-RAN deployment for future 6G and beyond. Although these values can be modified for edge cases and other future situations and are not a standard to follow. To prevent distortion in case of such scenarios, wideband carrier aggregation (such as $\geq$ 100 MHz) may require greater margins, but advanced precoding or clipping can be used to reduce the PAPR back-off margin. Also, power amplifier efficiency ($\eta_{\text{PA}}$) frequently deteriorates at mmWave frequencies because of linearity and thermal restrictions, although methods like envelope tracking and digital pre-distortion (DPD) can provide some compensation. Cooling, DC-DC conversion, and synchronization losses are taken into account by the overhead term $P_{\text{misc,RU}}$. Because these characteristics can be changed externally, future users or xApps will be able to adapt the model to changing 6G configurations or hardware variances.

\section{Integration of the RU Power Model into ns3-oran}
\label{sec:ru-integration}

To enable accurate simulation of energy consumption at the RU level in O-RAN, we implemented a modular and extensible RU power model within the \texttt{ns3-oran} framework. This subsection discusses the key software components introduced or extended, their interaction with existing modules such as PHY and MAC layers, and the overall integration workflow. The goal is to enable reproducibility, future customization, and evaluation of different RU hardware or energy-saving techniques. The following components were developed or adapted to support RU-level power modeling in line with O-RAN specifications:

\begin{itemize}
    \item \textbf{\texttt{CalculateRUCurrent()}}: A custom utility function that estimates RU current draw as a function of transmit power (\texttt{TxPower}) and configurable hardware parameters, including power amplifier (PA) efficiency, DC-DC and mains supply loss, and the number of transceivers.
    
    \item \textbf{\texttt{BasicEnergySourceHelper}}: Used to instantiate and configure energy sources for RU nodes. Each RU is assigned a predefined energy capacity (\SI{100000}{J} in our experiments).
    
    \item \textbf{\texttt{SimpleDeviceEnergyModel}}: Attached to each RU node to model the actual current draw dynamically. The current value is derived using the custom RU power model.
    
    \item \textbf{\texttt{BasicEnergyHarvesterHelper}} (optional): Supports future extensions involving renewable energy or dynamic energy harvesting models.
\end{itemize}

\subsection{Integration Workflow}

The integration process operates by passing the $P_\mathrm{tx}$ and other important parameters that are considered as constants. Afterwards, the RU power model calculates the expected current draw using:
    \begin{center}
        \texttt{double currentA = CalculateRUCurrent(txPower, ...);}
    \end{center}

\noindent This current is set for each RU node using:
    \begin{center}
        \texttt{energyModel->SetCurrentA(currentA);}
    \end{center}
The device energy model is then attached to the node’s energy source. During the simulation, \texttt{ns-3} tracks the remaining energy in each node, enabling computation of:
    \begin{itemize}
        \item \textbf{Energy Consumption:} \( E = E_\text{initial} - E_\text{remaining} \)
        \item \textbf{Energy Efficiency:} \( \eta = \frac{\text{Total Bits Received}}{E} \)
    \end{itemize}


\subsection{Interaction with PHY and MAC Layers}

The RU energy model operates in a non-intrusive manner by reading transmission power directly from the PHY layer (\texttt{LteEnbPhy}). It does not modify internal logic in PHY or MAC layers, preserving compatibility with other modules. This separation of concerns allows researchers to reuse this energy model across different MAC scheduling or handover configurations in \texttt{ns3-oran}.

\subsection{Software Architecture Overview}

\begin{figure}[htbp]
    \centering
    \includegraphics[scale=0.75]{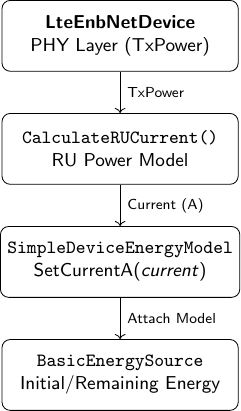}
    \caption{Software integration of RU power model into \texttt{ns3-oran}.}
    \vspace{-0.4cm}
    \label{fig:ru-power-arch}
\end{figure}

The integration is visually summarized in Figure~\ref{fig:ru-power-arch}, illustrating data flow from PHY-layer \texttt{TxPower} to current modeling and energy tracking.

\subsection{Advantages and Customizability}
The proposed integration offers the following advantages:
\begin{itemize}
    \item \textbf{Modularity}: Easily replaceable with alternate models such as DRX or ML-enhanced approaches.
    \item \textbf{O-RAN Aligned}: Focused specifically on RU-level energy modeling, decoupled from monolithic base station assumptions.
    \item \textbf{Extendable}: Can accommodate real-world hardware specs, dynamic transceiver activation, or energy-saving xApps.
    \item \textbf{Non-invasive}: Does not modify internal scheduling or PHY operations, ensuring compatibility with future \texttt{ns3-oran} releases.
\end{itemize}

This framework can serve as a foundational module for evaluating energy-aware orchestration, dynamic RU control, and xApp integration in future O-RAN research.


\section{Simulation and Results}
\label{sec:sim_results}
For the simulation setup, we have used the \texttt{ns-3.41} simulator with O-RAN extensions from \texttt{ns3-oran} \cite{ns3oran_github} to implement and evaluate the introduced RU power model. A two-cell LTE scenario is configured with UEs moving between the cells to create handover opportunities. A single Near-RT RIC hosts the xApps to manage control logic and trigger power adjustments. The RU power model is integrated at the PHY level of each eNB, using a custom logic based on the EARTH framework and extended to account for transceiver count and transmission dynamics. Simulation duration is fixed at 30 seconds to ensure consistency across all Tx power sweeps. Table~\ref{tab:sim-params} summarizes the simulation parameter used during the simulation. All relevant code and datasets are published to the GitHub Repository in \cite{ru_enery_ns3_oran}.


\begin{table}[htbp]
\centering
\caption{Simulation Parameters}
\label{tab:sim-params}
\resizebox{\linewidth}{!}{
\begin{tabular}{ll}
\toprule
\textbf{Parameter} & \textbf{Value} \\
\midrule
Simulator Version & \texttt{ns-3.41} \\
RAN Architecture & O-RAN (Near-RT RIC + xApps) \\
eNB Count & 2 \\
UE Count & 4 \\
Simulation Time & 30 seconds \\
UE Mobility & Constant velocity (1.5 m/s) \\
Handover Interval & 15 seconds \\
eNB Distance & 50 meters \\
Tx Power Sweep & 20--49 dBm \\
Power Model & Custom RU model (EARTH-inspired) \\
Transceiver Count & 64 per eNB \\
Voltage & 48 V \\
Current Calculation & Based on PA efficiency + fixed overhead \\
Traffic Pattern & UDP, On/Off (10 Mbps peak) \\
\bottomrule
\end{tabular}
}
\vspace{-0.5cm}
\end{table}

\begin{figure*}[htbp]
    \centering
    \begin{subfigure}{0.48\textwidth}
        \centering
        \includegraphics[scale= 0.45]{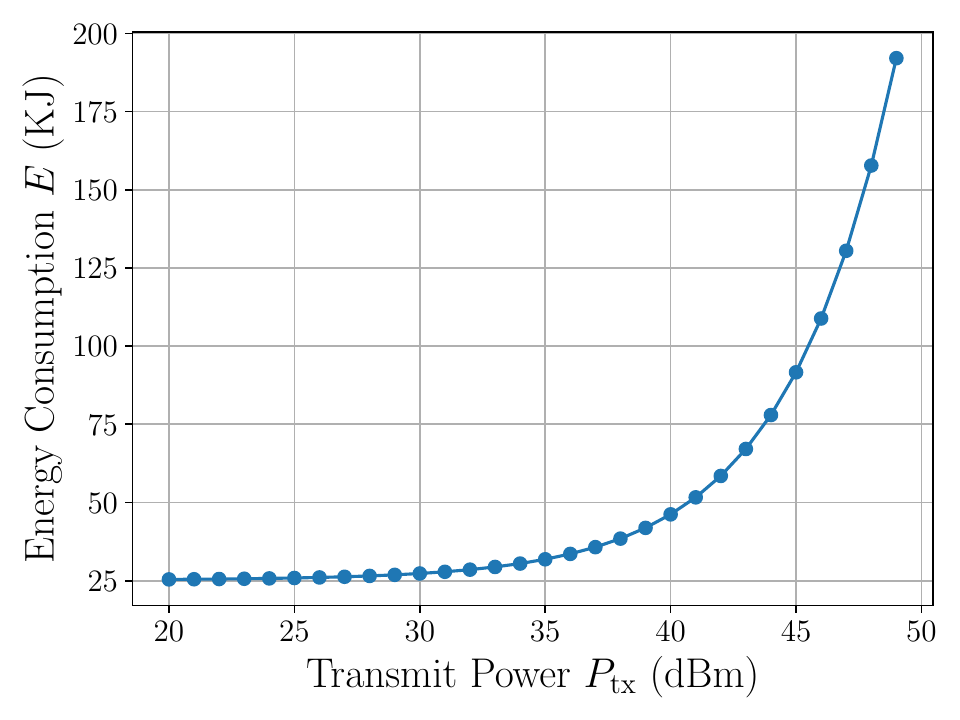}
        \caption{Energy consumption $E$ versus transmit power $P_\mathrm{tx}$.}
        \label{fig:energy-vs-txpower}
    \end{subfigure}
    \hfill
    \begin{subfigure}{0.48\textwidth}
        \centering
        \includegraphics[scale= 0.45]{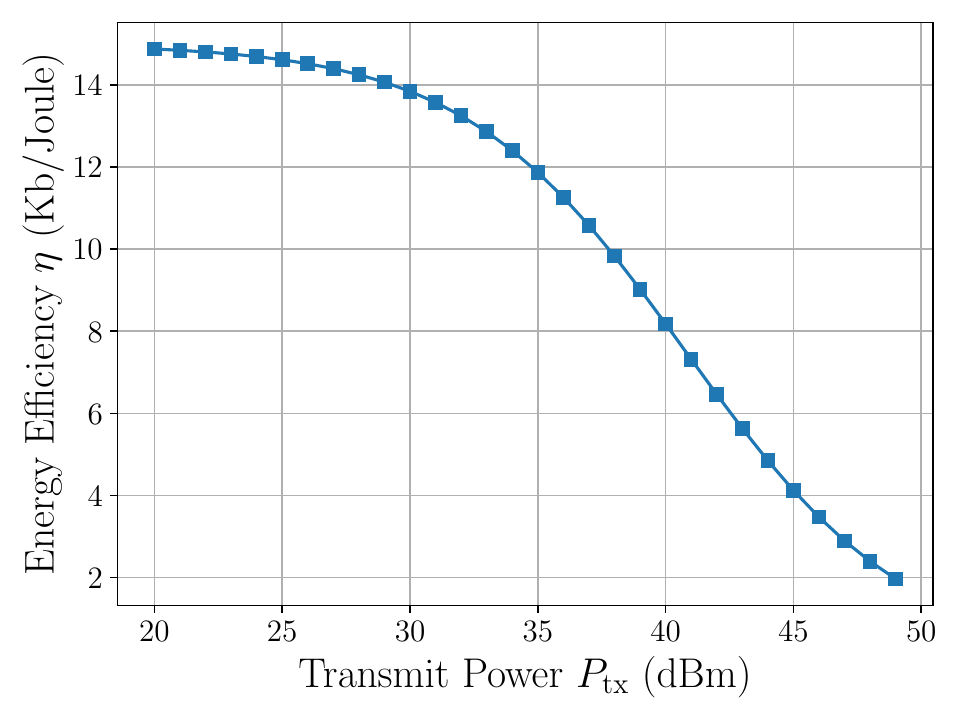}
        \caption{Energy efficiency $\eta$ versus transmit power $P_\mathrm{tx}$.}
        \label{fig:efficiency-vs-txpower}
    \end{subfigure}
    
    \medskip
    
    \begin{subfigure}{0.48\textwidth}
        \centering
        \includegraphics[scale= 0.45]{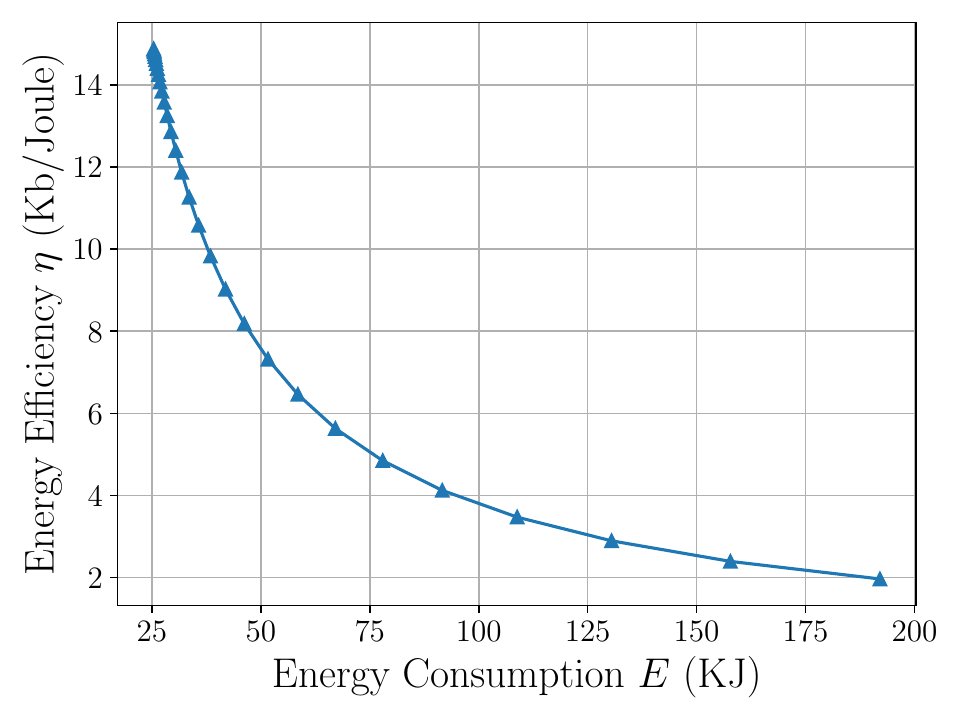}
        \caption{Energy efficiency $\eta$ versus energy consumption $E$.}
        \label{fig:efficiency-vs-energy}
    \end{subfigure}
    \hfill
    \begin{subfigure}{0.48\textwidth}
        \centering
        \includegraphics[scale= 0.35]{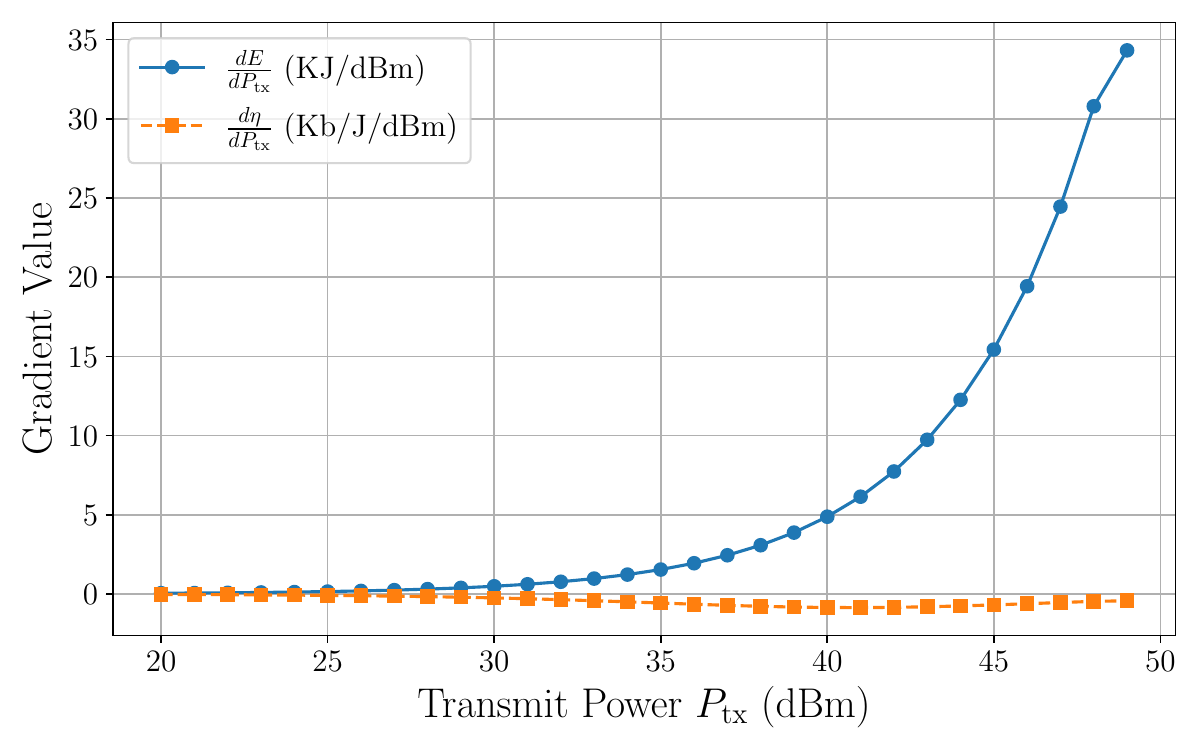}
        \caption{Gradient of energy efficiency $\eta$ and energy consumption $E$ versus transmit power $P_\mathrm{tx}$.}
        \label{fig:gradient-efficiency-energy-vs-txpower}
    \end{subfigure}
    
    \caption{Simulation results for the introduced RU Energy Model in \texttt{ns3-oran}.}
    \label{fig:combined-2x2}
    \vspace{-0.5cm}
\end{figure*}


Fig.~\ref{fig:energy-vs-txpower} illustrates the nonlinear increase in energy consumption $E$ as the transmit power $P_\mathrm{tx}$ rises. The energy model reflects realistic scaling behavior due to the power amplifier’s inefficiency and the inclusion of fixed overheads and transceiver counts in the introduced RU model. The upward trend is gradual at lower power levels but accelerates beyond 30 dBm, validating the quadratic characteristics embedded in the model. This behavior supports the model’s sensitivity to high-power configurations and underscores its utility in evaluating trade-offs in O-RAN deployments involving variable Tx power.


Fig.~\ref{fig:efficiency-vs-txpower} reveals an inverse relationship between energy efficiency $\eta$ (measured in Kilo-bits per joule) and Tx power $P_\mathrm{tx}$. The introduced model successfully captures the diminishing return in efficiency as transmission power $P_\mathrm{tx}$ increases, highlighting a steep drop beyond 30 dBm. Notably, the peak efficiency is observed near 20 dBm, marking it as a potential sweet spot for energy-optimal operation. This observation aligns with design principles in modern green RAN strategies and provides a quantitative basis for energy-aware scheduling in O-RAN xApps.


The energy consumed by the RUs and their corresponding energy efficiency is illustrated in Fig.~\ref{fig:efficiency-vs-energy}. The near-hyperbolic shape emphasizes that higher energy usage does not translate to better performance. The efficiency degrades sharply as more energy is consumed. This visualization reinforces the importance of finding an optimal Tx power $P_\mathrm{tx}$ configuration and demonstrates the practical impact of the RU model in evaluating performance-per-energy trade-offs across deployment scenarios.


Gradient analysis is presented in dual-plot Fig.~\ref{fig:gradient-efficiency-energy-vs-txpower} to evaluate how rapidly energy consumption and efficiency respond to changes in Tx power. The first derivative of energy ($\frac{dE}{dP_{\mathrm{tx}}}$) increases monotonically which confirms that non-linear scaling—likely quadratic or exponential—due to amplifier and supply losses. Conversely, the derivative of energy efficiency ($\frac{d\eta}{dP_{\mathrm{tx}}}$) exhibits steep negative slopes in higher Tx power regimes, emphasizing how efficiency degradation accelerates. This further justifies the model’s structure and its effectiveness in detecting inefficiencies at scale.


\section{Conclusion}
\label{sec:conclusion}
This work addresses the lack of fine-grained power modeling in existing open-source network simulation frameworks like ns-3 by introducing a simulator-integrated RU energy model specifically designed for O-RAN. The model allows for accurate prediction of RU power in active and sleep phases by extending the EARTH model and adding O-RAN-specific elements such as disaggregation, beamforming, and transceiver scaling. The introduced model's accuracy, scalability, and use for assessing energy efficiency trends and control tactics are demonstrated via extensive validation throughout Tx power sweeps. Future research on energy-saving xApps, dynamic RU control, and sustainable 5G/6G RAN orchestration will be built upon this architecture.


%




\ifCLASSOPTIONcaptionsoff
  \newpage
\fi




\begin{thebibliography}{10}


\bibitem{wadud2023conflict}
A.~Wadud, F.~Golpayegani, and N.~Afraz, ``Conflict Management in the Near-RT-RIC of Open RAN: A Game Theoretic Approach,'' in {\em 2023 IEEE International Conferences on Internet of Things (iThings) and IEEE Green Computing \& Communications (GreenCom) and IEEE Cyber, Physical \& Social Computing (CPSCom) and IEEE Smart Data (SmartData) and IEEE Congress on Cybermatics (Cybermatics)}, pp.~479--486, IEEE, 2023.


\bibitem{wadud2024qacm}
A.~Wadud, F.~Golpayegani, and N.~Afraz, ``QACM: QoS-Aware xApp Conflict Mitigation in Open RAN,'' {\em IEEE Transactions on Green Communications and Networking}, vol.~8, no.~3, pp.~978--993, 2024.


\bibitem{ns3oran_github}
U.S. National Institute of Standards and Technology (NIST), ``ns3-oran,'' {\em GitHub Repository}, 2025. [Online]. Available: \url{https://github.com/usnistgov/ns3-oran}. [Accessed: February 15, 2025].

\bibitem{boyapati2010energy}
H.~K. Boyapati, R.~V.~R. Kumar, and S.~Chakrabarti, ``Energy consumption assessment in LTE baseband of eNodeB and guidelines for green baseband subsystems,'' in {\em Proceedings of the 2010 Annual IEEE India Conference (INDICON)}, pp.~1--6, IEEE, 2010.

\bibitem{liang2024energy}
X.~Liang, Q.~Wang, A.~Al-Tahmeesschi, S.~B. Chetty, D.~Grace, and H.~Ahmadi, ``Energy Consumption of Machine Learning Enhanced Open RAN: A Comprehensive Review,'' {\em IEEE Access}, vol.~12, pp.~81889--81902, 2024.

\bibitem{auer2011energy}
G.~Auer, V.~Giannini, C.~Desset, I.~Godor, P.~Skillermark, M.~Olsson, M.~A. Imran, D.~Sabella, M.~J. Gonzalez, O.~Blume, and A.~Fehske, ``How much energy is needed to run a wireless network?,'' {\em IEEE Wireless Communications}, vol.~18, no.~5, pp.~40--49, 2011.

\bibitem{abubakar2023energy}
A.~I. Abubakar, O.~Onireti, Y.~Sambo, L.~Zhang, G.~K. Ragesh, and M.~A. Imran, ``Energy efficiency of open radio access network: A survey,'' in {\em 2023 IEEE 97th Vehicular Technology Conference (VTC2023-Spring)}, pp.~1--7, IEEE, 2023.

\bibitem{oran_potential_2025}
O.-R. Alliance, ``Potential Energy Savings Features in O-RAN,'' {\em White Paper}, no.~2025-01, 2025. [Online]. Available: \url{https://mediastorage.o-ran.org/white-papers/O-RAN.SuFG.Potential%20Energy%20Savings%20Features%20in%20O-RAN%20white%20paper%202025-01.pdf}.

\bibitem{akman2024energy}
A.~Akman, P.~Oliver, M.~Jones, P.~Tehrani, M.~Hoffmann, and J.~Li, ``Energy Saving and Traffic Steering Use Case and Testing by O-RAN RIC xApp/rApp Multi-vendor Interoperability,'' in {\em 2024 IEEE 100th Vehicular Technology Conference (VTC2024-Fall)}, pp.~1--6, IEEE, 2024.


\bibitem{rimedo_whitepaper_2023}
M.~Hoffmann and M.~Dryjanski, ``The O-RAN Whitepaper 2023: Energy Efficiency in O-RAN,'' {\em Rimedo Labs}, 2023. [Online]. Available: \url{https://rimedolabs.com/blog/the-oran-whitepaper-2023-energy-efficiency-in-oran/}.

\bibitem{intel_vran_2023}
Intel and Mavenir, ``A Holistic Study of Power Consumption and Energy Savings Strategies for Open vRAN Systems,'' Technical Report, 2023. [Online]. Available: \url{https://builders.intel.com/docs/networkbuilders/a-holistic-study-of-power-consumption-and-energy-savings-strategies-for-open-vran-systems-1676628842.pdf}.

\bibitem{wang2023minimizing}
L.~Wang, J.~Zhou, M.~Ma, and X.~Niu, ``Minimizing energy consumption of IoT devices for O-RAN based IoT systems,'' {\em Energy Reports}, vol.~9, pp.~379--388, 2023.


\bibitem{hoffmann2024energy}
M.~Hoffmann and M.~Dryja{\'n}ski, ``Energy Efficiency in Open RAN: RF Channel Reconfiguration Use Case,'' {\em IEEE Access}, 2024.


\bibitem{nafea2021study}
H.~B. Nafea, M.~M. Sallam, and F.~W. Zaki, ``Study of DRX sleep mode performance on virtual base station energy saving in 5G networks,'' {\em Wireless Personal Communications}, vol.~118, no.~4, pp.~3251--3270, 2021.


\bibitem{ru_enery_ns3_oran}
A.~Wadud, ``RU-Energy-Model-ns3-oran,'' 2025. [Online]. Available: \url{https://github.com/dewanwadud1/RU-Energy-Model-ns3-oran}. [Accessed: April 14, 2025].

\end{thebibliography}
%
\bibliographystyle{ieeetr}
\end{document}